\documentclass[twocolumn]{aa}  
\usepackage[utf8]{inputenc}
\PassOptionsToPackage{hyphens}{url}
\usepackage[english]{babel}
\usepackage{natbib}
\usepackage{graphicx}
\usepackage{amsmath}
\usepackage[varg]{txfonts}
\usepackage{setspace}
\usepackage{epstopdf}
\usepackage{tabularx}
\usepackage{ulem}
\usepackage[colorlinks=true, linkcolor=blue, citecolor=blue, urlcolor=blue, breaklinks=true]{hyperref}
\usepackage{booktabs}

\newcommand{\fsequa}{\, \, \, .}
\newcommand{\comequa}{\, \, \, ,}

\newcommand{\Msol}{\mathrm{M}_{\odot}}
\newcommand{\Rsol}{\mathrm{R}_{\odot}}
\newcommand{\Lsol}{\mathrm{L}_{\odot}}

\newcommand{\yr}{\mathrm{yr}}
\newcommand{\Gyr}{\mathrm{Gyr}}

\newcommand*\diff{\mathop{}\!\mathrm{d}}
\newcommand{\bef}{\begin{figure}}
\newcommand{\eef}{\end{figure}}

\begin{document}

\institute{
Hamburger Sternwarte, Universit\"at Hamburg, Gojenbergsweg 112, 21029 Hamburg, Germany \label{inst1}
\and
Departamento de Astronom\'ia, Facultad Ciencias F\`isicas y Matem\`aticas, Universidad de Concepci\'on, Av. Esteban Iturra s/n Barrio Universitario, Casilla 160-C, Concepci\'on, Chile \label{inst2}}

\title{The physics of the Applegate mechanism: Eclipsing time variations from magnetic activity}

\titlerunning{Lanza model for PCEBs}

\author{M. Völschow\inst{\ref{inst1}}, D. R. G. Schleicher\inst{\ref{inst2}}, R. Banerjee\inst{\ref{inst1}} and J. H. M. M. Schmitt\inst{\ref{inst1}}}

\authorrunning{Völschow et al.}

\offprints{{\tt mvoelschow@hs.uni-hamburg.de}}

\abstract{Since its proposal in 1992, the Applegate mechanism has been discussed as a potential intrinsical mechanism to explain transit timing variations in various kinds of close binary systems. Most analytical arguments presented so far focused on the energetic feasibility of the mechanism, while applying rather crude one- or two-zone prescriptions to describe the exchange of angular momentum within the star. In this paper, we present the most detailed approach to date to describe the physics giving rise to the modulation period from kinetic and magnetic fluctuations. Assuming moderate levels of stellar parameter fluctuations, we find that the resulting binary period variations are one or two orders of magnitude lower than the observed values in RS-CVn like systems, supporting the conclusion of existing theoretical work that the Applegate mechanism may not suffice to produce the observed variations in these systems. The most promising Applegate candidates are low-mass post-common-envelope binaries (PCEBs) with binary separations $\lesssim 1~\Rsol$ and secondary masses in the range of $0.30~\Msol$ and $0.36~\Msol$.}


\maketitle 

\section{Introduction}
\label{sec:intro}
Precise timing measurements in close binaries routinely reveal variations in the eclipse timings. While for binaries with very short periods less than three hours gravitational wave emission is the dominating means of angular momentum loss, magnetic braking can account for such effects in binaries with longer periods \citep[see, e.g.][]{Kraft1962, Faulkner1971, Verbunt1981, Parsons2013}. A subclass of close binaries, mainly RS Canum Venaticorum (RS CVn) and post-common envelope binary (PCEB) systems, feature cyclic or nearly-periodic orbital period variations on timescales of a few years to decades incompatible with classical gravitational wave or magnetic braking models \citep[see, e.g.][]{Lanza1998, Brinkworth2006, Zorotovic2013}. For some systems, planetary companions may explain these variations \citep[see, e.g.][]{Qian2011, Beuermann2013, Nasiroglu2017, Han2017}, while some of the planetary solutions turned out to be dynamically unstable \citep{Horner2013} or have been proven wrong observationally \citep{Hardy2015}. On the other hand, the period variations in other systems such as QS~Vir are still not well understood \citep{Parsons2010}.\\
Based on and inspired by earlier work by \citet{Matese1983} and \citet{Applegate1987}, \citet{Applegate1992} proposed a new mechanism to explain cyclic orbital period variations in close binaries. The author assumes a time-dependent gravitational quadrupole moment modulated by the stellar activity cycle. Given a constant orbital angular momentum, an increasing quadrupole moment results in a stronger gravitational field and finally in a decreasing orbital radius and increasing orbital velocity, which can be observed as a reduction of the binary period.\\
To calculate the required energy as well as the expected quadrupole changes, \citet{Applegate1992} considered a thin shell rotating in a point mass potential representing the rest of the star. The author assumed that the torque necessary to perform the angular momentum exchange is provided by subsurface magnetic fields of a few $\mathrm{kG}$. Given such a field, luminosity variations on the $10~\%$ level and angular velocity variations on the $1~\%$ level are expected, leading to binary period variations compatible with the observed values in RS CVn systems ($\Delta P / P \sim 10^{-5}$). \citet{Applegate1992} made a number of testable predictions which include significant period luminosity changes due to temperature variations and a 1:1 relation between the modulation period and the stellar activity cycle. Both the former and the latter have been observed in systems such as \textit{CG Cygni}, accompanied by orbital period variations \citep[][]{Hall1991}, while luminosity variations can be observed by studying photospheric temperatures \citep[see, e.g.][]{Gray1994}.\\ 
\citet{Lanza1998} generalized the magnetohydrodynamic aspects of the pioneering work of \citet{Applegate1992}, included the effects of internal magnetic fields and elaborated on the connection between the Applegate mechanism and different types of dynamo models, emphasizing that a careful study of the Applegate mechanism may allow to distinguish between them. The improved \citet{Lanza1998} model is based on the momentum balance equation and assumes a uniform ratio of plasma pressure to magnetic pressure over the star. Expansion of the gravitational potential in spherical harmonics allowed \citet{Lanza1998} to calculate the quadrupole potential at the surface of the star, finding that the angular velocity changes required to produce a certain amount of period variation are reduced by a factor of 2 compared to the original \citet{Applegate1992} approach.\\
Furthermore, \citet{Lanza1998} found that the driver of the period variations is the magnetic dynamo of the active component, which can effectively transform kinetic energy from nuclear reactions into magnetic energy and is more likely to be an $\alpha^{2}\Omega$ dynamo, introducing a further observational aspect. \citet{Lanza1999} stressed that the original \citet{Applegate1992} ansatz may systematically overestimate the energies required to power the quadrupole moment variations as the exchanges between kinetic and magnetic energy are cyclic and partially reversible processes. Therefore, the authors concluded that the original model only provides upper limits for the required energy.\\
\citet{Lanza2005} revisited the original work by \citet{Applegate1992} and introduced an entirely new approach. Assuming that the angular velocity of the active component is only a function of the distance from its rotation axis, \citet{Lanza2005} considered angular momentum redistributions not only between two layers but rather general radial transport and re-distribution modes within the convection zone. Imposing a strictly adiabatic convection zone and neglecting density perturbations, the author considered the equations of mass continuity and angular momentum conservation within a mean-field framework. Solving the angular momentum equation, \citet{Lanza2005} calculated individual angular momentum redistribution modes from which he derived both the amplitude of the quadrupole moment change and the kinetic energy dissipated during the redistribution processes. Contrary to prior works, \citet{Lanza2005} found that the power required to drive the observed levels of binary period variation in HR 1099 exceeds the luminosity provided by the active component by one or two orders of magnitude. \citet{Lanza2006} extended the \citet{Lanza2005} approach to angular velocity distributions that are functions of both radius and latitude, finding that the Applegate effect is still not a viable option in the case of HR 1099 and RS CVn systems in general.\\ 
Another route has been taken by \citet{Brinkworth2006} which generalized the original \citet{Applegate1992} thin-shell ansatz to angular momentum exchanges between a finite core and a finite shell, including the core's back-reaction to oblateness changes of the surrounding shell. \citet{Voelschow2016} extended the \citet{Brinkworth2006} model to realistic stellar density profiles, derived analytic expressions to estimate the energy required to produce a certain level of period variation in a given system, and applied their full model to a set of 16 close binary systems, including eleven PCEBs. In line with previous works, the authors found that the Applegate effect cannot uniquely explain period variations in close binaries, and only four of the 16 systems were identified as potential Applegate candidates. Most recently, \citet{Navarrete2018} employed the analytic two-zone model by \citet{Voelschow2016} to investigate how rotation impacts on the energetic feasibility of the Applegate effect and on which scale the activity cycle matches the observed modulation period, noting that the systems with the highest rotation rates are the most likely Applegate candidates. However, one critical simplification of both the \citet{Brinkworth2006} and \citet{Voelschow2016} models is that the authors set the core-shell transition where it minimizes the energy necessary to power the Applegate effect.\\
In this paper, we extend the \citet{Lanza2006} model by assuming a time-dependent magnetic field, velocity field fluctuations and magnetic field fluctuations in the convection zone from which we can explicitly calculate the temporal evolution of the dissipated energy as well as the binary period variation. In addition, we consider superpositions of angular momentum redistribution modes instead of calculating the quadrupole moment changes caused by individual modes as previously done by \citet{Lanza2005} and \citet{Lanza2006}. For the sake of consistency, we first apply our model to HR 1099 and use the system as an illustrative example for the basic predictions we make, before we extend our analysis to PCEB systems and identify the most likely Applegate mechanism candidates. 
\section{Model description}
\label{sec:model}
The formalism we employed is based on \citet{Lanza2006}. Sec.~\ref{sec:setup} and \ref{sec:ang_mom} closely follow the author's description, but our presentation puts an emphasis on a more algorithmic description. Starting from sec.~\ref{sec:source_term} we extend on the crucial aspects of the calculation by making an ansatz for the stellar parameter fluctuations, explicitly solving the temporal part of the underlying differential equation (see sec.~\ref{sec:ang_mom}), elaborating on the evolution of the mechanism and presenting an approach how to describe superpositions of angular momentum redistribution modes.
\subsection{Setup}
\label{sec:setup}
\citet{Lanza2006} assumes an inertial reference frame originating at the barycenter of the active component star, with the z-axis pointing into the direction of the stellar rotation axis. In line with \citet{Lanza2006}, we employ a spherical coordinate system with $r$ being the distance from the origin, colatitude $\theta$ measured from the north pole and azimuthal angle $\phi$ and impose that all variables are independent of $\phi$. The hydrodynamics of the turbulent convection zone are described employing a mean-field approach $\vec{V} = \bar{\vec{v}} + \vec{v}'$ with mean velocity $\bar{\vec{v}}$ and mean value fluctuation $\vec{v}'$. Furthermore, we neglect the impact of meridional circulations inside the star by assuming that the mean velocity field arises purely from stellar rotation \citep{Lanza2006}.
The equation for the angular velocity $\omega = v_\mathrm{\phi}/(r \sin \theta)$ reads \citep[see, e.g.][]{Lanza2005, Lanza2006}
\begin{equation}
\label{eq:angmom}
\begin{split}
&\quad \frac{\partial \omega}{\partial t} - \frac{1}{\rho r^4} \frac{\partial}{\partial r} \left( r^4 \eta _\mathrm{t} \frac{\partial \omega}{\partial r} \right) - \frac{\eta _\mathrm{t}}{\rho r^2} \frac{1}{(1-\mu^2)} \frac{\partial}{\partial \mu} \left( (1-\mu^2)^2 \frac{\partial \omega}{\partial \mu} \right) = \\ 
&\quad = S(r, \mu, t) \fsequa
\end{split}
\end{equation}
Here, $\eta _\mathrm{t} = \eta _\mathrm{t}(r)$ is the turbulent dynamical viscosity, $\mu = \cos \theta$. $S(r, \mu, t)$ is a source term that controls the temporal evolution of the angular momentum redistribution and will be specified in sec.~\ref{sec:source_term}.
Eq.~(\ref{eq:angmom}) is solved assuming a stress-free boundary condition
\begin{equation}
\label{eq:omega_condition}
\left( \frac{\partial \omega}{\partial r} \right) _\mathrm{r_\mathrm{b}, R} = 0 \comequa
\end{equation}
$r_\mathrm{b}$ denotes the base of the stellar convection zone, $R$ is the radius of the star \citep{Lanza2006}.
\subsection{The angular momentum equation}
\subsubsection{Model framework}
\label{sec:ang_mom}
Introducing the state of rigid rotation $\Omega _\mathrm{0}$, the angular velocity of the star can be written as
\begin{equation}
\label{eq:omega_star}
\Omega (r, \mu, t) = \Omega _\mathrm{0} + \omega (r, \mu, t) \comequa
\end{equation}
where $\omega (r, \mu, t)$ describes the deviation from rigid rotation \citep{Lanza2006}.
According to \citet{Lanza2006}, solutions of the angular velocity equation~(\ref{eq:angmom}) have the form
\begin{equation}
\label{eq:omega_n}
\omega (r, \mu, t) = \sum \limits _{n=0} ^{\infty} \alpha _{n} (t) \zeta _{n} (r) P_{n} ^{(1,1)} (\mu) \comequa
\end{equation}
where $P_{n} ^{(1,1)}$ are Jacobian polynomials and the index $n$ will be refered to as angular order. Further, following \citet{Lanza2006} the source term $S(r, \mu, t)$ can be developed into:
\begin{equation}
\label{eq:s_series}
S (r, \mu, t) = \sum \limits _{n=0} ^{\infty} \beta _{n} (t) \zeta _{n} (r) P_{n} ^{(1,1)} (\mu) \fsequa
\end{equation}
Based on the angular momentum conservation equation and using the properties of the Jacobian polynomials, \citet{Lanza2006} derived a partial differential equation for the $\alpha _{n}$, $\beta _{n}$ and $\zeta _{n}$ functions that can be separated into two equations which can be solved independently from each other, namely
\begin{equation}
\label{eq:alpha}
\frac{\diff \alpha _{n}(t)}{\diff t} - \beta _{n}(t) + \alpha_{n}(t)  \lambda _{n} = 0 \comequa
\end{equation}
which we refer to as temporal equation as well as
\begin{equation}
\label{eq:zeta}
\frac{1}{\rho(r) r^4} \frac{\diff}{\diff r} (r^4 \eta_\mathrm{t}(r) \zeta' _{n}(r)) -n(n+3) \frac{\eta_\mathrm{t}(r)}{\rho(r) r^2} \zeta_{n}(r) + \lambda _{n} \zeta _{n}(r) = 0 \comequa
\end{equation}
which we shall refer to as radial equation, and $\lambda _n$ are eigenvalues of eq.~(\ref{eq:zeta}). The radial eq.~(\ref{eq:zeta}) together with the boundary conditions (\ref{eq:omega_condition}) define a regular Sturm-Liouville problem in the interval $[r_\mathrm{b},R]$ for $r_\mathrm{b} > 0$ \citep{Lanza2006}.
\subsubsection{Properties of the radial equation}
For any given $n$ an infinite number of eigenvalues exist. Therefore, \citet{Lanza2006} introduced a new index $k$ (radial order) and denominated as $\lambda _{nk}$ the k-th eigenvalue of the n-th radial equation. The first eigenvalue $\lambda _{00}$ is zero and the associated eigenfunction $\zeta _{00}$ vanishes and all eigenvalues are positive \citep[see][]{Lanza2006}.\\
Accounting for the radial order $k$, eq.~(\ref{eq:omega_n}) can be recast into
\begin{equation}
\label{eq:omega_nk}
\omega (r, \mu, t) = \sum \limits _{k} \sum \limits _{n=0} ^{\infty} \alpha _{nk} (t) \zeta _{nk} (r) P_{n} ^{(1,1)} (\mu) \comequa
\end{equation}
and the radial equation eq.~(\ref{eq:zeta}) now reads
\begin{equation}
\label{eq:zeta_nk}
\frac{1}{\rho r^4} \frac{\diff}{\diff r} (r^4 \eta_\mathrm{t} \zeta' _{nk}) - n(n+3) \frac{\eta_\mathrm{t}}{\rho r^2} \zeta_{nk} + \lambda _{nk} \zeta _{nk} = 0 \fsequa
\end{equation}
Imposing $|\omega| \ll \Omega _\mathrm{0}$ (cf. eq.~\ref{eq:omega_star}) and employing a linear approximation for the variation of the gravitational quadrupole moment (see sec.~\ref{sec:delta_q}), we only have to consider the cases $n=0$ and $n=2$ \citep{Lanza2006}.
Strictly speaking, in the case of a convective zone that spans the entire star ($r_\mathrm{b} = 0$) the regular Sturm-Liouville problem turns into a singular Sturm-Liouville problem. The boundary condition at the lower end of the domain must be replaced by a regularity condition and a different treatment of Eq.~(\ref{eq:zeta_nk}) is required to deal with singularities at $r=0$. However, the stellar structure models we employ (see sec.~\ref{sec:results}) all start at some radial coordinate $r>0$ which ensures $r_\mathrm{b} > 0$ even for fully-convective stars.
\subsubsection{Solving the radial equation}
The most essential input to solve the radial equations is a stellar structure model that provides the mass distribution within the star $M(r)$, the luminosity $L(r)$, the density $\rho(r)$, the temperature $T(r)$ and the mean molecular weight $m _\mathrm{\mu}(r)$. Based on this input, we can derive the gravitational acceleration
\begin{equation}
g(r) = \frac{G M(r)}{r^2} \comequa
\end{equation}
and the convective velocity employing mixing length theory $u_\mathrm{c}$
\begin{equation}
u_\mathrm{c} (r) = \left( \frac{\alpha_\mathrm{ml} \, L(r)}{40 \, \pi \, r^2 \, \rho (r)} \right)^{1/3} \comequa
\end{equation}
where we will use a mixing length parameter $\alpha_\mathrm{ml} = 1.5$ \citep{Lanza2005}. The pressure scale height $h_\mathrm{p}$ is given by
\begin{equation}
h_\mathrm{p} (r) =  \frac{k_\mathrm{B} \, T(r)}{m_\mathrm{H} \, m _\mathrm{\mu} (r) \, g (r)} \comequa
\end{equation}
with Boltzmann's constant $k_\mathrm{B}$ and hydrogen mass $m_\mathrm{H}$ \citep{Lanza2005}. Finally, we can calculate the turbulent viscosity from
\begin{equation}
\eta_\mathrm{t} (r) = \frac{1}{3} \, \rho (r) \, \alpha_\mathrm{ml} \, u_\mathrm{c}(r) \, h_\mathrm{p}(r) \fsequa
\end{equation}
The stress-free boundary condition specified by eq.~(\ref{eq:omega_condition}) implies $\zeta _{nk} ' (r) = 0$ at $r = r_\mathrm{b}$ and $r = R$ as boundary conditions for $\zeta _{nk}$. We assume that $\zeta _{nk}$ vanishes outside $[r_\mathrm{b},R]$, i.e. angular velocity variations only occur in the convection zone. Intermediate stellar structure values are calculated by a linear interpolation scheme.\\
Solutions for $\zeta _{0k}$ and $\zeta _{2k}$ are computed via a shooting method \citep[see, e.g.][]{Fehlberg1987, Press1992}. While the boundary conditions imply $\zeta _{nk} ' (r_\mathrm{b}) = 0$, the eigenfunction normalization at the bottom of the convective zone $\zeta _{nk} (r_\mathrm{b})$ can be regarded as a free parameter of the model only restricted by $\zeta _{nk} (r_\mathrm{b}) \ll \Omega _\mathrm{0}$. The impact of this parameter will be addressed in section \ref{sec:results}.\\
\citet{Lanza2006} adopted $\zeta _{nk} (r_\mathrm{b}) = 0.01 \, \Omega _\mathrm{0}$ throughout his calculations, i.e. normalized individual angular momentum redistribution modes to some level. In contrast, we solve the full temporal equation to find solutions for the $\alpha _{nk}(t)$ and $\beta _{nk}(t)$ functions allowing us to calculate the evolution of the redistribution processes and consider superpositions of the elemental redistribution functions $\zeta _{nk}(r)$.
\subsection{The temporal equation}
\subsubsection{Framework}
\label{sec:source_term}
Following \citet{Lanza2006}, solutions for the temporal equation~(\ref{eq:alpha}) are
\begin{equation}
\label{eq:alpha_nk}
\alpha _{nk} (t) = \exp (- \lambda _{nk} t) \int \limits _{0} ^{t} \beta _{nk} (t') \exp (\lambda _{nk} t') \diff t' + \alpha _{nk} (0) \comequa
\end{equation}
where $\beta _{nk}$ is calculated from 
\begin{equation}
\label{eq:beta_nk}
\beta _{nk} = E _{nk} \int \limits _{r _\mathrm{b}} ^{R} \int \limits _{-1} ^{1} \rho r^4 S(r, \mu, t) \zeta _{nk} P_{n} ^{(1,1)} (1-\mu^2) \diff \mu \diff r \fsequa
\end{equation}
$E _{nk}$ is a normalization constant\footnote{The expression for the normalization constant can be derived by multiplying eq.~(\ref{eq:s_series}) with $\rho r^4 \zeta _{nk} P_{n} ^{(1,1)} (1-\mu^2)$, integrating as $\int \limits _{r _\mathrm{b}} ^{R} \int \limits _{-1} ^{1}  \diff \mu \diff r$ and making use of the orthogonality of $\zeta _{nk}$ for a fixed $n$ with respect to the weight function $\rho r^4$, and the orthogonality of the Jacobian polynomials.} given by
\begin{equation}
\label{eq:e_nk_start}
E _{nk}^{-1} = \int \limits _{r_\mathrm{b}} ^{R} \rho r^4 \zeta _{nk}^2 \diff r \, \int \limits _{-1} ^{1} P_{n} ^{(1,1)} P_{n} ^{(1,1)} (1-\mu^2) \diff \mu  \fsequa
\end{equation}
The integral involving the Jacobian polynomials simplifies to
\begin{equation}
\int \limits _{-1} ^{1} P_{n} ^{(1,1)} P_{n} ^{(1,1)} (1-\mu^2) \diff \mu = \frac{8 \, (n+1)}{(2n+3)(n+2)} \fsequa
\end{equation}
In a linear approximation for the quadrupole moment variation, we only have to consider the angular orders $n=0$ and $n=2$ (see sec.~\ref{sec:delta_q}) for which the term on the right hand side can be evaluated as $4/3$ for $n=0$ and $6/7$ for $n=2$.\\
The source term $S$ in eq.~(\ref{eq:angmom}) describes angular momentum transfer by Reynolds stresses and magnetic torques and takes the form
\begin{equation}
S(r, \mu, t) = - \frac{\mathrm{div} \, \vec{\tau}}{\rho r^2 (1-\mu^2)} \fsequa
\end{equation}
The vector $\vec{\tau}_i$ has the components:
\begin{equation}
\tau_i = r \sin \theta \left( \vec{\Lambda} _{i \phi} + \frac{1}{\tilde{\mu}} (B_i B_\phi + \vec{M_{i \phi}}) \right) \fsequa
\end{equation}
$\vec{\Lambda} _{i \phi} = \rho \, \overline{v_i' v_\phi'}$ is the Reynolds stress tensor, $\tilde{\mu}$ is the magnetic permeability, $\vec{B}$ is the mean magnetic field and $\vec{M}_{i \phi} = \overline{B_i' B_\phi'}$ is the Maxwell stress tensor \citep[]{Lanza2005}. Finally, we arrive at the expression
\begin{equation}
\beta_{nk} = -E_{nk} \int \limits _{-1} ^{1} \int \limits _{r_b} ^{R} r^{2} \mathrm{div} \, \vec{\tau} \, \zeta _{nk} P_{n} ^{(1,1)} \diff \mu \diff r
\end{equation}
to calculate the $\beta_{nk}$ functions from which we can then calculate the $\alpha_{nk}$ functions necessary to describe the temporal evolution of the star's angular momentum distribution.\\
\subsubsection{Stellar parameter fluctuations}
As the Applegate mechanism is linked to and triggered by magnetic activity, we impose a phase factor of the form
\begin{equation}
f(t) = \sin \left( \omega _\mathrm{act} t \right)
\end{equation} 
with $\omega _\mathrm{act} = 2 \pi / P_\mathrm{act}$ to describe the fluctuations of all involved quantities where $P_\mathrm{act}$ is the activity cycle period \citep[see, e.g.][]{Ruediger2002}. For the magnetic field, we assume a simple azimuthal structure with $B_r = 0$, $B_\theta = 0$ and
\begin{equation}
B_\phi (t) = B_\mathrm{surf} \, \sin \left( \omega _\mathrm{act} t \right)
\end{equation}
accompanied by magnetic field fluctuations of the form
\begin{equation}
 B'_{i}(r,t) = A_\mathrm{B} \, B_\mathrm{surf} \, \sin \left( \omega _\mathrm{act} t \right)
\end{equation}
and velocity field fluctuations in the convection zone 
\begin{equation}
 v'_{i}(r,t) = A_\mathrm{v} \, u_\mathrm{c}(r) \, \sin \left( \omega _\mathrm{act} t \right)
\end{equation}
where $A_\mathrm{B}$ and $A_\mathrm{v}$ are amplification coefficients and $B_\mathrm{surf}$ is the surface magnetic field amplitude. Assuming these equations describe proper long-term ensemble means and imposing that both $B'_{i}B'_{j}$ and $v'_{i}v'_{j}$ are independent statistics for $i \neq j$, we can make an ansatz for the Reynolds tensor
\begin{equation}
 \Lambda _{i \phi}(r,t) = \rho (r) A_\mathrm{v}^2 u_\mathrm{c}^2 (r) \, \sin^{2}(\omega _\mathrm{act} t)
\end{equation}
as well as the Maxwell stress tensor:
\begin{equation}
 M _{i \phi}(r,t) = A_\mathrm{B}^2 B_\mathrm{surf}^2 \, \sin^{2}(\omega _\mathrm{act} t) \fsequa
\end{equation}
The impact of a constant phase lag $\Delta \phi$ between the phase factor of different coordinates will be addressed in Sec.~\ref{sec:results_period_modulation}. We note that the stellar parameter fluctuation description we adopt for our model does not come from dynamo theory or hydrodynamic calculation, but is rather a simplified ad-hoc approach to generate cyclic Maxwell and Reynolds stresses.
\subsubsection{Solution of the temporal equation}
We can now evaluate the source term $S(r, \mu, t)$. For a vector field $\vec{F}(r,\theta,\phi)$ in spherical coordinates, the divergence takes the form
\begin{equation}
\mathrm{div} \, \vec{F} = \left( \frac{1}{r^{2}} \frac{\partial}{\partial r} (r^{2} F_{r}), \frac{1}{r \sin \theta} \frac{\partial}{\partial \theta}(\sin \theta  \, F_\theta), \frac{1}{r \, \sin \theta} \frac{\partial}{\partial \phi} (F_\phi) \right)^\mathrm{T} \fsequa
\end{equation}
For our given magnetic field configuration and fluctuation setup, the divergence of the vector $\vec \tau$ can be calculated as
\begin{equation}
\mathrm{div} \, \vec{\tau} = \sqrt{1-\mu^2} \, \frac{1}{r^{2}} \frac{\partial}{\partial r} \left[ r^{3} \, \tilde{B}_{r\phi}(r,t) \right] + 2 \mu \tilde{B}_{\theta\phi}
\end{equation}
where we defined
\begin{equation}
\label{eq:b_tilde}
 \tilde{B}_{i\phi}(r,t) = \Lambda _{i \phi} (r,t) + \frac{B_i (r,t) B_\phi (r,t)}{\tilde{\mu}} + \frac{M_{i \phi}(r,t)}{\tilde{\mu}} \fsequa
\end{equation}
Using this, the function $\beta _{nk}(t)$ can be evaluated as
\begin{equation}
\beta _{nk} = - E_{nk} \, \int \limits _{r_\mathrm{b}} ^{R} \zeta _{nk}(r) \frac{\partial}{\partial r} \left[ r^{3} \tilde{B}_{r\phi}(r,t) \right] \diff r \, \int \limits _{-1} ^{1} P_{n} ^{(1,1)} \sqrt{1-\mu^2} \diff \mu \fsequa
\end{equation}
Note that the second term from $\mathrm{div} \, \tau$ vanishes because of the symmetry of the $\mu$ integral. The remaining Jacobian polynomial integral can be evaluated analytically. First, we substitute $\mu = \sin u$ and use the identity $\sin^2 u + \cos^2 u = 1$ which yields
\begin{equation}
\label{eq:jacobi_sqrt}
\begin{split}
&\quad \int \limits _{-1} ^{1} P_{n} ^{(1,1)} \sqrt{1-\mu^2} \diff \mu = \\ 
&\quad \int \limits _{-\pi/2} ^{\pi/2} P_{n} ^{(1,1)} (\sin u) \diff u - \int \limits _{-\pi/2} ^{\pi/2} P_{n} ^{(1,1)} (\sin u) \sin^2 (u) \diff u \fsequa
\end{split}
\end{equation}
Next, $P_{n} ^{(1,1)}(\mu)$ can be written as
\begin{equation}
P_{n} ^{(1,1)}(\mu) = \frac{1}{n+2} \sum \limits _{m=0} ^{n} \frac{(n+m+2)!}{m! (n-m)! (m+1)! 2^m} (\mu-1)^m
\end{equation}
and for $(\mu-1)^m$ we have
\begin{equation}
 (\mu-1)^m = \sum \limits _{k=0} ^{m} (-1)^k \frac{m!}{k! (m-k)!} \mu^{m-k} \fsequa
\end{equation}
Using
\begin{equation}
 \int \limits _{-\pi/2} ^{\pi/2} \sin^q (u) \diff u = \pi \frac{2^{-q-1} q! \left((-1)^q + 1\right) }{(q/2)!^2}
\end{equation}
the integrals in eq.~(\ref{eq:jacobi_sqrt}) are given by
\begin{equation}
\label{eq:jacobi_analytic1}
\begin{split}
&\quad \int \limits _{-\pi/2} ^{\pi/2} P_{n} ^{(1,1)} (\sin u) \diff u = \\
&\quad \frac{\pi}{2 (n+2)} \sum \limits _{m=0} ^n \frac{(n+m+2)!}{(n-m)!(m+1)! 2^{2m}} \sum \limits _{k=0} ^m \frac{(-2)^k}{k!} \frac{\left( (-1)^{m-k}+1 \right)}{((m-k)/2)!^2} 
\end{split}
\end{equation}
as well as
\begin{equation}
\label{eq:jacobi_analytic2}
\begin{split}
&\quad \int \limits _{-\pi/2} ^{\pi/2} P_{n} ^{(1,1)} (\sin u) \sin^2 u \diff u = \\
&\quad \frac{\pi}{8 (n+2)} \sum \limits _{m=0} ^n \frac{(n+m+2)!}{(n-m)!(m+1)! 2^{2m}} \cdot \\
&\quad \sum \limits _{k=0} ^m \frac{(-2)^k}{k!} \frac{\left( (-1)^{m-k+2}+1 \right) (m-k+1)(m-k+2)}{((m-k+2)/2)!^2}
\end{split}
\end{equation}
Because of the symmetry of the integrand, the integral vanishes for odd values of $n$ implying $\beta_{nk} = 0$ and $\alpha_{nk} = 0$. Using eq.~(\ref{eq:jacobi_analytic1}) and eq.~(\ref{eq:jacobi_analytic2}) allows us to explicitly calculate the Jacobian integral eq.~(\ref{eq:jacobi_sqrt}) (see tab.~\ref{tab:results_jacobi}).
\begin{table}
\caption{Results for the Jacobian integral as given by eq.~(\ref{eq:jacobi_sqrt}).}
\begin{center}
\begin{tabular}{cc}
\toprule
Order $n$ & Result	\\
\midrule
               0   &  $\pi / 2$ \\
               2   &  $3 \pi / 32$ \\
               4   &  $5 \pi / 128$ \\
               6   &  $175 \pi / 8192$ \\
               8   &  $441 \pi / 32768$ \\
               10   &  $4851 \pi / 524288$ \\
\bottomrule
\end{tabular}
\label{tab:results_jacobi}
\end{center}
\end{table}
In order to normalize and calculate $\alpha _{nk}$, we have to make use of the initial conditions which correspond to the state of rigid rotation. Formally, we have
\begin{equation}
 \Omega (r,0,0) = \Omega _\mathrm{0}
\end{equation}
which implies
\begin{equation}
 \omega (r,0,0) = 0
\end{equation}
and can be satisfied by demanding $\alpha _{nk} (0) = 0$.
\subsection{Energy dissipation}
According to \citet{Lanza2006}, the variation of the rotational kinetic energy can be calculated from
\begin{equation}
\Delta \mathcal{T} _{nk} = \frac{8 \pi \, (n+1)}{(2n+3)(n+2)} \alpha ^2 _{nk} \int \limits _{r_\mathrm{b}} ^{R} \rho r^4 \zeta _{nk}^2 \diff r \comequa
\end{equation}
with a kinetic energy dissipation rate $P_\mathrm{diss}$ given by
\begin{equation}
P_\mathrm{diss} = -2 \, \sum \limits _k \sum \limits _n \lambda _{nk} \, \Delta \mathcal{T} _{nk} \fsequa
\end{equation}
\subsection{Gravitational quadrupole moment variation}
\label{sec:delta_q}
According to \citet{Ulrich1981} and \citet{Lanza2006}, the variation of the quadrupole moment potential $\Delta \Phi _{12}$ can be calculated by solving
\begin{equation}
\label{eq:dp12}
\begin{split}
&\quad \frac{\partial^2 (\Delta \Phi _{12})}{\partial r^2} = - \frac{2}{r} \frac{\partial (\Delta \Phi _{12})}{\partial r} + \frac{6}{r^2}(\Delta \Phi _{12}) + \\ 
&\quad \frac{4 \pi r^2}{M(r)} \left[ \left( \frac{\diff \rho}{\diff r} \right) (\Delta \Phi _{12}) - \frac{\partial}{\partial r} \left( r^2 \, \rho \, b_2 \right)  - r \, \rho \, a_2 \right] \comequa
\end{split}
\end{equation}
where $M(r)$ is the integrated star mass up to a given radius $r$. The \citet{Lebovitz1970} coefficients $a_2$ and $b_2$ are calculated from the radial eigenfunctions $\zeta _{nk}$
\begin{equation}
\label{eq:lebovitz_a2}
a_2 = \Omega _0 \sum \limits _{k} \frac{12}{7} \alpha _{2k} \zeta _{2k} - \frac{4}{3} \alpha _{0k} \zeta _{0k}
\end{equation}
and
\begin{equation}
\label{eq:lebovitz_b2}
b_2 = \Omega _0 \sum \limits _{k} \frac{2}{3} \alpha _{0k} \zeta _{0k} + \frac{4}{7} \alpha _{2k} \zeta _{2k} \fsequa
\end{equation}
Solutions for eq.~(\ref{eq:dp12}) must verify the condition
\begin{equation}
\label{eq:dp12_condition}
\Delta \Phi _{12}' + 3 \Delta \Phi _{12} / R = 0
\end{equation}
to match the outer gravitational potential \citep{Lanza2006}.
\subsubsection{Solving the quadrupole moment equation}
In order to solve eq.~(\ref{eq:dp12}) with the boundary condition eq.~(\ref{eq:dp12_condition}), we perform a shooting-method search and convert this boundary value problem to an initial value problem. The initial conditions are given by $\Delta \Phi _{12}(r) = C \, r^2$ and $\Delta \Phi _{12}'(r) = 2 \, C \, r$ for $r \rightarrow 0$. By performing subsequent integrations of the quadrupole moment equation~(\ref{eq:dp12}) for varying trial constants $C$, we find a trial constant $C$ for which condition~(\ref{eq:dp12_condition}) holds.\\
From $\Delta \Phi _{12}(R)$, we can calculate the quadrupole moment variation $\Delta Q$ via \citep{Lanza2006}
\begin{equation}
\Delta Q = -\frac{R^3 \, \Delta \Phi _{12}(R)}{3 \, G} \comequa
\end{equation}
giving a relative orbital period variation of \citep{Applegate1992}
\begin{equation}
\frac{\Delta P}{P} = -9 \, \frac{\Delta Q}{M \, a^2} \comequa
\end{equation}
where $a$ is the semi-major axis of the binary and $M$ the mass of the active component.
\section{Results}
\label{sec:results}
We applied the formalism described in Sec.~\ref{sec:model} to an EZAMS\footnote{\url{http://www.astro.wisc.edu/~townsend/static.php?ref=ez-web}} model of HR1099\footnote{We chose HR1099 for the sake of comparability with previous works by \citet{Lanza2005} and \citet{Lanza2006} upon which our models is based on.} ($M = 1.3~\Msol$, solar metallicity, timestep 426, $t = 4.48~\Gyr$, $R \sim 4.05~\Rsol$, $L = 8.62~\Lsol$, $\Omega _0 = 2.569 \cdot 10^{-5}~\mathrm{s}^{-1}$, depth of the convection zone $r_\mathrm{b}/R = 0.1968$), which provides the necessary stellar data: radial mass profile $M(r)$, luminosity profile $L(r)$, density profile $\rho(r)$, temperature profile $T(r)$ and mean molecular weight $m _\mathrm{\mu}(r)$, as well as the base of the convection zone $r_b$. Using this, we can calculate derived quantities such as the radial gravitational acceleration profile $g(r)$, convective velocity $u_\mathrm{c}(r)$, pressure scale height $h_\mathrm{p}(r)$, and turbulent viscosity $\eta_\mathrm{t}(r)$ for which we employ mixing length theory with a mixing length parameter $\alpha _\mathrm{ml} = 1.5$. We plot the radial profiles of $u_\mathrm{c}(r)$, $h_\mathrm{p}(r)$ and $\eta_\mathrm{t}(r)$ in Fig.~\ref{fig:star}.\\
\bef
\centering
\resizebox{\hsize}{!}{\includegraphics{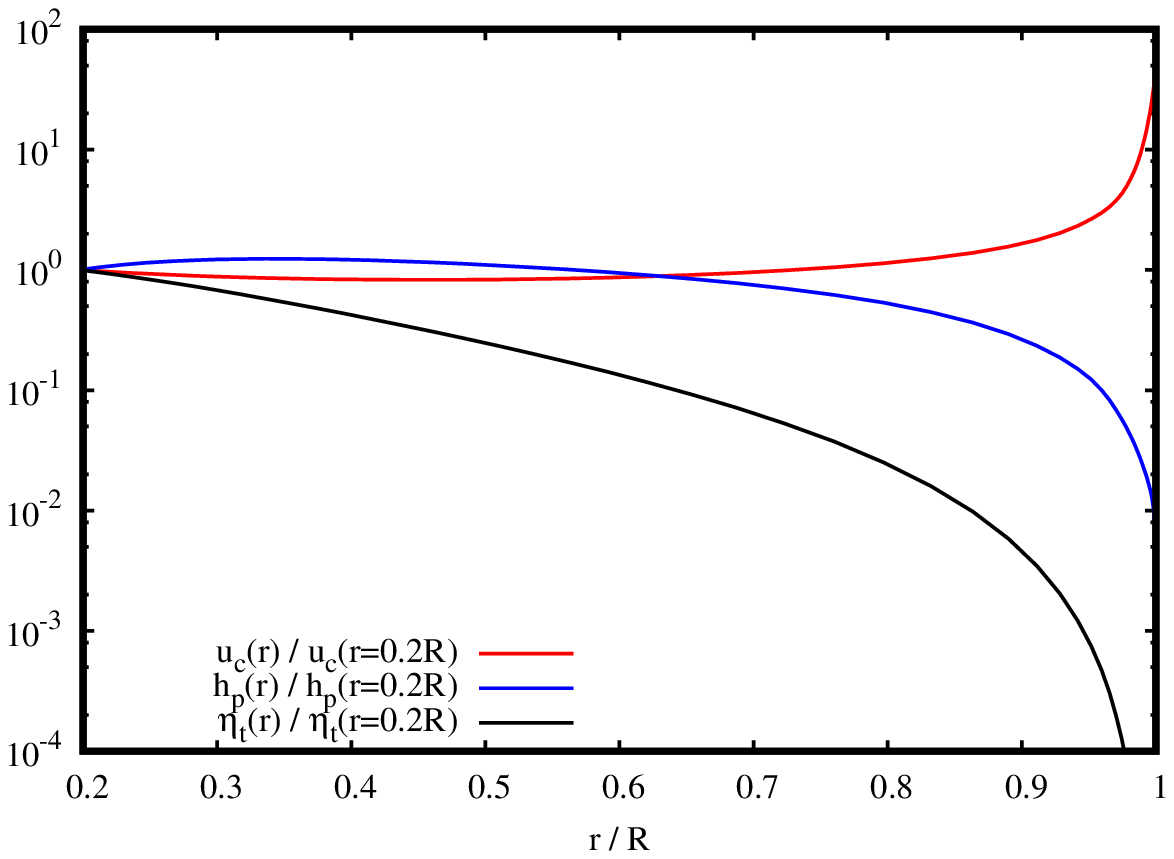}}
\caption{Convective velocity profile $u_\mathrm{c}$, pressure scale height $h_\mathrm{p}$ and turbulent dynamical viscosity $\eta _\mathrm{t}$ calculated using an EZAMS model of HR1099 and mixing length theory.}
\label{fig:star}
\eef
For the surface magnetic flux density, we assume $B_\mathrm{surf} = 1~\mathrm{kG}$. Details on the calculation of the Maxwell stress tensor and Reynolds stress tensor are described in Sec.~\ref{sec:model}. We adopt $A_v = A_B = 0.1$ as relative oscillation amplitudes for the velocity field and magnetic field inside the star.
\subsection{Analytical approach}
\label{sec:analytical}
In order to understand the temporal evolution of the Applegate effect in the model presented here, we shall start with an analytical approach towards the most fundamental quantities involved in the mechanism. For a given set of functions that describe the oscillation of stellar parameters, the mechanism is controlled by the functions $\beta _{nk}$ and $\alpha _{nk}$ which can be calculated from the source term $S(r, \mu, t)$ - which is in turn controlled by the function $\tilde{B}_{i\phi}(r,t)$ as given by eq.~(\ref{eq:b_tilde}). $\tilde{B}_{i\phi}(r,t)$ consists of three terms:
\begin{itemize}
\item A magnetic field term $B_i B_\phi(r,t)$ controlling the mean magnetic field, 
\item the Reynolds tensor $\Lambda _{i\phi}(r,t)$ describing the velocity field fluctuations,
\item the Maxwell tensor $M _{i\phi}(r,t)$ which describes the magnetic field fluctuations.

\end{itemize}
Let $f(t)$ be a periodic function with period $P_\mathrm{act}$ (implying $f(t+P_\mathrm{act}) = f(t)$) that describes the temporal variation of the stellar parameters involved in the Applegate mechanism, i.e. $B_i' (r,t) = B_i(r) \, f(t)$, and $v_i' (r,t) = u_c(r) \, f(t)$. Further, we employ a cyclic purely azimuthal magnetic field $B_\phi (t) = B_\mathrm{surf} \, f(t)$ with constant amplitude $B_\mathrm{surf}$. We impose that all these quantities are tightly coupled and oscillate in phase. Under such conditions, all terms in $\tilde{B}_{i\phi}(r,t)$ have the same time dependence implying $\tilde{B}(r,t) \propto f(t)^2$ for fixed r. Substituting into eq.~(\ref{eq:beta_nk}) yields $\beta _{nk} \propto f(t)^2$. Given this result, we can evaluate eq.~(\ref{eq:alpha_nk}): Following \citet{Lanza2006} and employing that the first eigenvalues are typically associated with angular momentum redistribution modes on timescales shorter than the activity cycle of the star, we have $\alpha _{nk} = \beta _{nk} / \lambda _{nk}$ which yields $\alpha _{nk} \propto f(t)^2$.\\
Finally, we arrive at
\begin{equation}
P_\mathrm{diss} \propto f(t)^4 \fsequa
\end{equation}
The main insight of this exercise is the change in periodicity: in case of sinusoidal variations with period $P_\mathrm{act}$ in the magnetic field and the velocity field fluctuations, $\alpha _{nk}$, $\beta _{nk}$ and $P_\mathrm{diss}$ oscillate with half of the activity period to the leading order\footnote{Using the multiple-angle formula $\cos 4x = 1 - 8 \, \sin ^2 x + 8 \, \sin ^4 x$ we see that $f(t)^2$ and $f(t)^4$ oscillate with the same amplitude and frequency, but are accompanied by a smaller oscillation with 4 times the frequency and $1/8$ the amplitude.} which is the observed binary modulation period $P_\mathrm{mod}$ as the quadrupole moment variation closely follows the dissipated energy (see following subsections).
Another interesting result can be found by calculating the dimensionless quantity
\begin{equation}
\frac{\tilde{\mu} \Lambda}{M} = \tilde{\mu} \frac{\int \limits _\mathrm{r_\mathrm{b}} ^\mathrm{R} A_v^2 u_c^2 \rho \diff r}{\int \limits _\mathrm{r_\mathrm{b}} ^\mathrm{R} A_B^2 B_\mathrm{surf}^2 \diff r} \comequa
\end{equation}
which compares the Reynolds tensor integral with the Maxwell tensor integral, quantifying the relative strength of the velocity field fluctuation amplitude versus the magnetic field fluctuation amplitude. Given $A_v = A_B = 0.1$, a surface magnetic field of $B_\mathrm{surf} = 1~\mathrm{kG}$, and our EZAMS model of HR 1099, we have $\tilde{\mu} \Lambda / M \gg 1$, implying that the velocity fluctuations dominate over the magnetic field fluctuations. The same holds for the zero-age main sequence stars investigated in the parameter study in sec.~\ref{sec:mass_ps}. As a result, the choice of $B_\mathrm{surf}$ and $A_B$ does not significantly alter our results.
\subsection{Eigenvalues and radial eigenfunctions}
\label{sec:eigenvalues}
\begin{table}
\caption{Summary of the first ten eigenvalues of the radial equation for $n=0$ and $n=2$.}
\begin{center}
\begin{tabular}{ccc}
\toprule
Order	$k$ & $\lambda _{0k} / \ s^{-1}$ & $\lambda _{2k} / \ s^{-1}$	\\
\midrule
0 &  0 & 3.54E-8 \\
1 & 5.12E-8 & 1.17E-7 \\
2 & 1.39E-7 & 2.27E-7 \\
3 & 2.64E-7 & 3.63E-7 \\
4 & 4.28E-7 & 5.27E-7 \\
5 & 6.31E-7 & 7.26E-7 \\
6 & 8.73E-7 & 9.64E-7 \\
7 & 1.16E-6 & 1.24E-6 \\
8 & 1.48E-6 & 1.56E-6 \\
9 & 1.84E-6 & 1.92E-6 \\
10 &  2.24E-6 & 2.32E-6 \\
\bottomrule
\end{tabular}
\label{tab:results_lambda}
\end{center}
\end{table}
Fig.~\ref{fig:zeta0k} and fig.~\ref{fig:zeta2k} show the first four eigenfunctions $\zeta _{0k}(r)$ and $\zeta _{2k}(r)$ associated with the first four eigenvalues of $\lambda _{0k}$ and $\lambda _{2k}$ (see tab.~\ref{tab:results_lambda}). A radial eigenfunction of order $k$ has $k$ nodes in $[r_\mathrm{b},R]$ \citep[see, e.g.][]{Lanza2006}.
\bef
\centering
\resizebox{\hsize}{!}{\includegraphics{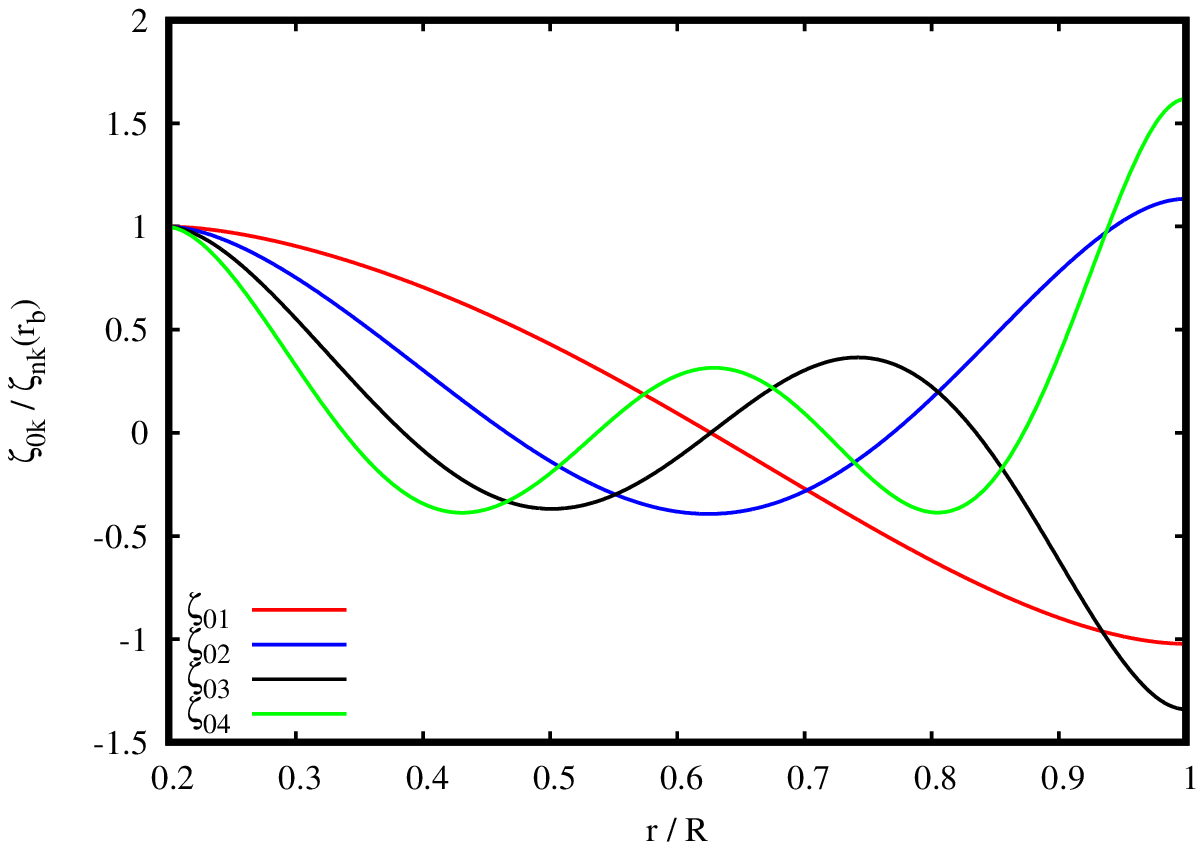}}
\caption{Eigenfunctions $\zeta _{0k}$ associated with the first four eigenvalues $\lambda _{0k}$ calculated for HR1099 (see Sec.~\ref{sec:eigenvalues}). These functions represent elemental angular momentum redistribution modes.}
\label{fig:zeta0k}
\eef
\bef
\centering
\resizebox{\hsize}{!}{\includegraphics{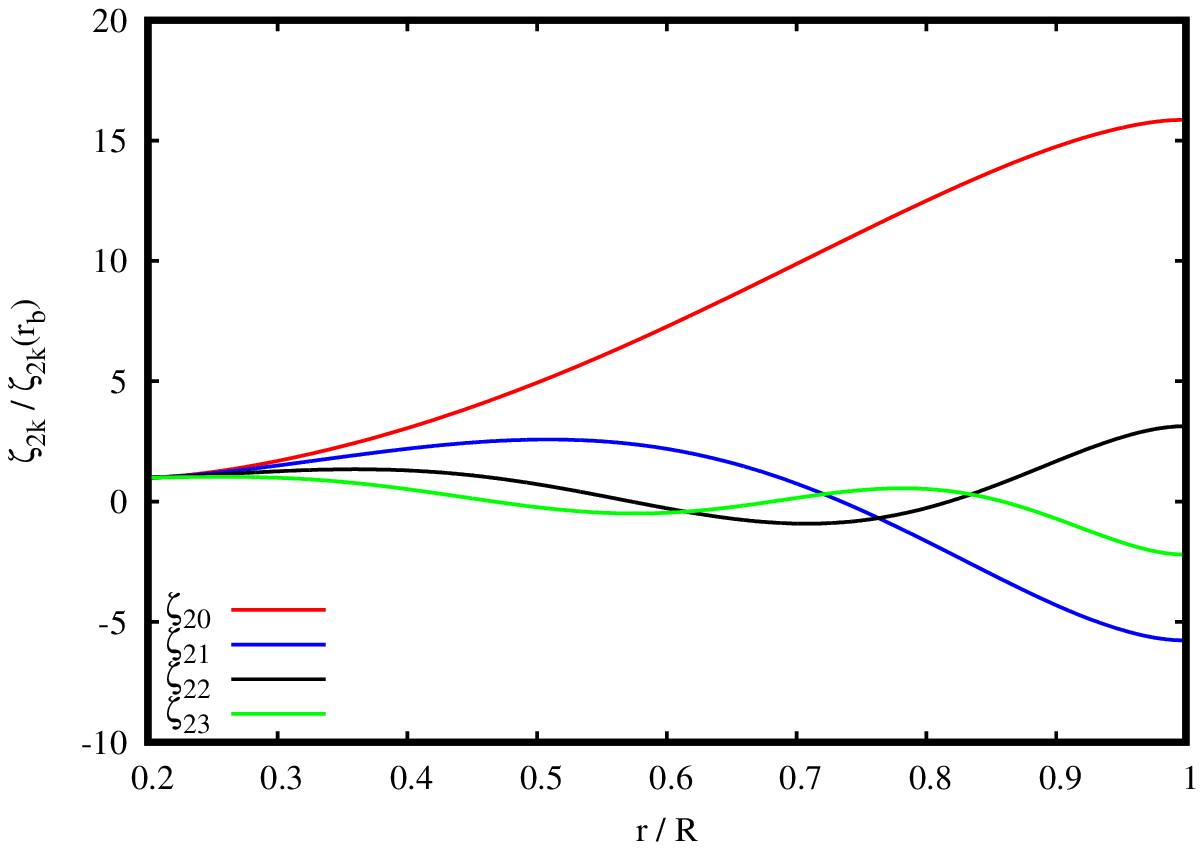}}
\caption{Eigenfunctions $\zeta _{2k}$ associated with the first four eigenvalues $\lambda _{2k}$ calculated for HR1099 (see Sec.~\ref{sec:eigenvalues}). These functions represent elemental angular momentum redistribution modes.}
\label{fig:zeta2k}
\eef
\subsection{Eigenfunction normalization and k cutoff}
\label{sec:normalization}
For a fixed $r_b$ and assuming that all radial eigenfunction $\zeta_{nk}$ share one normalization parameter $\zeta _{nk} (r_\mathrm{b})$, we find that the scaling relations $\omega \propto \zeta _{nk} (r_\mathrm{b})$ and $P_\mathrm{diss} \propto \zeta _{nk} (r_\mathrm{b}) ^2$ hold. Consider $\zeta _{nk} (r) = \zeta _{nk} (r_\mathrm{b}) \, \xi _{nk}(r)$ with dimensionless functions $\xi _{nk}(r)$. Plugging into eq.~(\ref{eq:zeta_nk}) and making use of the linearity of differentials yields
\begin{equation}
\frac{1}{\rho r^4} \frac{\diff}{\diff r} (r^4 \eta_\mathrm{t} \xi' _{nk}) - n(n+3) \frac{\eta_\mathrm{t}}{\rho r^2} \xi_{nk} + \lambda _{nk} \xi _{nk} = 0 \comequa
\end{equation}
which is equivalent to eq.~(\ref{eq:zeta_nk}). This implies that for a fixed value of $r_\mathrm{b}$, $n$ and $k$, solutions for $\zeta _{nk} (r)$ with varying normalizations $\zeta _{nk} (r_\mathrm{b})$ are multiples of one another. In particular, the eigenvalues $\lambda _{nk}$ are constant. With this, we obtain for the dissipated energy
\begin{equation}
\Delta \mathcal{T} _{nk} = \zeta^2 _{nk} (r_\mathrm{b}) \, \frac{8 \pi \, (n+1)}{(2n+3)(n+2)} \alpha ^2 _{nk} \int \limits _{r_\mathrm{b}} ^{R} \rho r^4 \xi _{nk}^2 \diff r
\end{equation}
and the kinetic energy dissipation rate becomes
\begin{equation}
\label{eq:p_diss}
P_\mathrm{diss} = -2 \, \sum \limits _k \sum \limits _n \lambda _{nk} \, \Delta \mathcal{T} _{nk} \propto \zeta^2 _{nk} (r_\mathrm{b}) \fsequa
\end{equation}
as the eigenvalues $\lambda _{nk}$ are independent from the normalization chosen. In an analogous way, we can write
\begin{equation}
\omega (r, \mu, t) = \zeta _{nk} (r_\mathrm{b}) \sum \limits _{n=0} ^{\infty} \sum \limits _k \alpha _{nk} (t) \xi _{nk} (r) P_{n} ^{(1,1)} (\mu)
\end{equation}
for the angular velocity variation inside the star which indeed implies $\omega (r, \mu, t) \propto \zeta _{nk} (r_\mathrm{b})$ because the functions $\alpha _{nk}$ are invariant under variations of $\zeta _{nk} (r_\mathrm{b})$.\\
As we investigate the question whether the Applegate effect - if operative - can explain the observed period variations in close binaries, we choose the following normalization strategy: First, we start with $\zeta _{nk} (r_\mathrm{b}) = 1$ and calculate the eigenvalues $\lambda _{nk}$ and the functions $\zeta _{nk}$, $\beta _{nk}$ and $\alpha_{nk}$ for $n=0$ and $n=2$ up to some maximum order $k_\mathrm{max}$. The maximum order is determined by the eigenvalues which have the dimension of inverse time and are associated with angular momentum redistributions on a timescale of $1/\lambda_{nk}$ \citep[see][]{Lanza2005, Lanza2006}. For increasing orders of $n$ and $k$, these timescales shrink gradually smaller until they fall below the typical travel time a soundwave takes to cross the convection zone which we calculate from
\begin{equation}
\label{eq:soundwave_travel}
t_\mathrm{c} = \frac{(R-r_\mathrm{b})^2}{\int \limits _{r_\mathrm{b}} ^{R} c_\mathrm{s} \diff r} 
\end{equation}
where the speed of sound is
\begin{equation}
c_\mathrm{s} = \sqrt{ 5 \, P / 3 \, \rho }
\end{equation}
assuming monoatomic gas.
From this point, we therefore assume that higher orders are not able to contribute to the global angular momentum redistribution process. Depending on the details of the stellar interior, typical k cutoffs range between 50 for evolved subgiants and >100 for red dwarfs.
Given a k cutoff and the eigenfunctions $\zeta _{nk}$ (starting with $\zeta _{nk} (r_\mathrm{b}) = 1$), $\beta _{nk}$ as well as $\alpha _{nk}$, we calculate the total energy dissipation $P_\mathrm{diss}(t)$ and locate its maximum $P_\mathrm{diss,max}$. Given that, we adjust the eigenfunction normalization factor $\zeta _{nk} (r_\mathrm{b})$ in order to have a maximum dissipated power equal to $A_\mathrm{P} \, L_\mathrm{HR1099}$. Using this normalization, the kinetic energy dissipation rate is limited to some fraction $A_\mathrm{P}$ of the stellar power output $L_\mathrm{Hr1099}$ which is typically of the order of $10~\%$, and we can finally calculate the corresponding angular velocity variations inside the star.
\subsection{Angular velocity variation and energy dissipation}
The dissipated energy is calculated from eq.~(\ref{eq:p_diss}) up to some radial order $k$. For $k = 48$, the timescale associated with the respective angular momentum redistribution mode is shorter than the calculated soundwave travel time. We fix the kinetic energy fluctuation limit to $A_\mathrm{P} = 0.1$ as described in the previous subsection. 
Once we found $\alpha_{nk}$ and $\zeta_{nk}$ and applied our normalization to limit the kinetic energy dissipation, we can calculate the effective angular velocity variation inside the star as both functions of time and stellar radius. In Fig.~\ref{fig:omega} we present the angular velocity changes inside the star for three different times between $t = 0.05 P_\mathrm{mod}$ and $t = 0.25 P_\mathrm{mod}$. Starting from the state of rigid rotation at $t = 0$, the angular velocity variation builds up continuously with peak fluctuations of a few per cent in the outermost layers of the star.
\bef
\centering
\resizebox{\hsize}{!}{\includegraphics{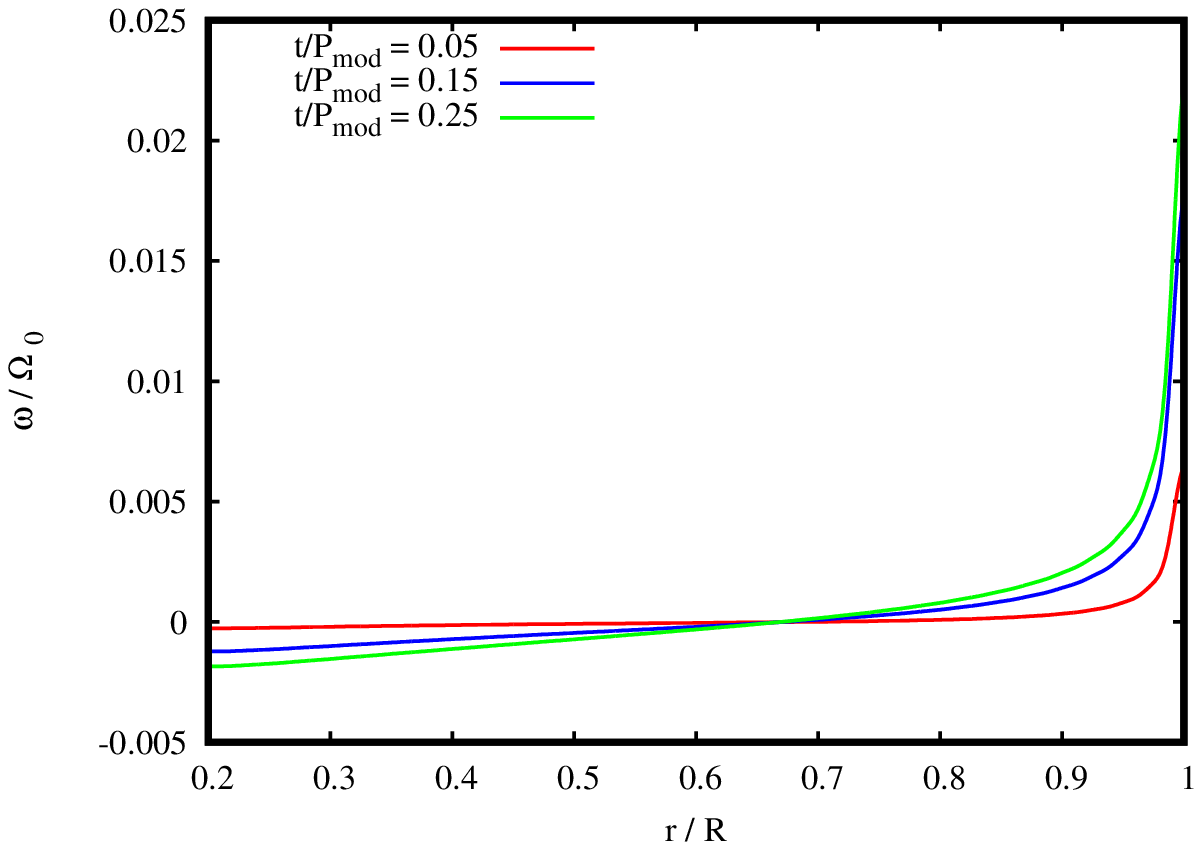}}
\caption{Evolution of the relative angular velocity variation inside HR 1099 computed for three different times.}
\label{fig:omega}
\eef
\subsection{Period modulation}
\label{sec:results_period_modulation}
In Fig.~\ref{fig:deltaP} we plot the calculated period variation in HR 1099 for $B_\mathrm{surf} = 1~\mathrm{kG}$, a cycle length of $P_\mathrm{act} = 2 \, P_\mathrm{mod}$, a magnetic field fluctuation amplitude of $A_B = 0.1$ and a velocity field fluctuation amplitude $A_V = 0.1$ for three different cases of phase lag $\Delta \phi$ between the radial and the azimuthal velocity and magnetic field fluctuations. The dissipated energy is limited to $A_P = 0.1$ of the star's luminosity. For $\Delta \phi = 0$, the oscillation is purely negative, i.e. the quadrupole moment changes have a positive sign. On the other hand, for $\Delta \phi = \pi / 2$, the period modulation oscillates between the positive and the negative regime indicative of a cyclic transition between an oblate and a prolate state.
\bef
\centering
\resizebox{\hsize}{!}{\includegraphics{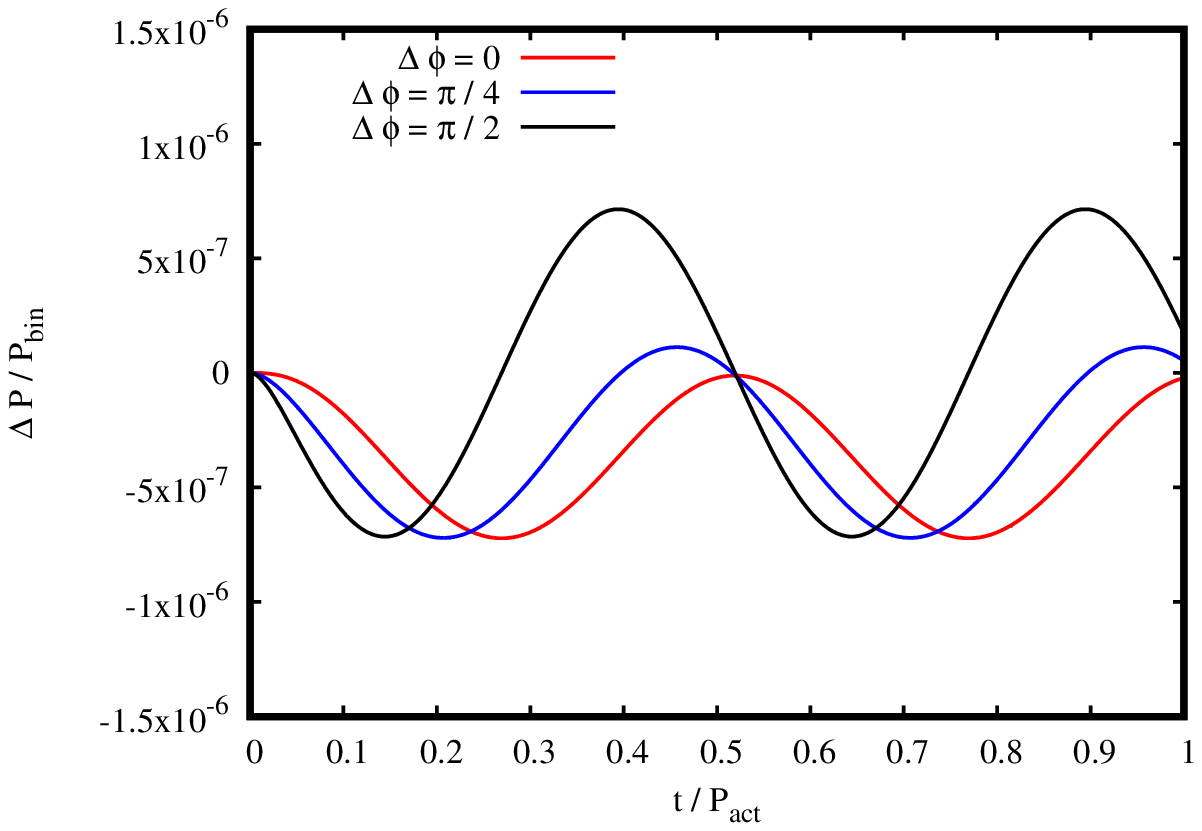}}
\caption{Binary period variation amplitude calculated for a surface magnetic field of $1~\mathrm{kG}$, a cycle length of $P_\mathrm{act} = 2 \, P_\mathrm{mod}$, a magnetic field fluctuation amplitude of $A_B = 0.1$ and a velocity field fluctuation amplitude $A_V = 0.1$ for three different cases of phase shift $\Delta \phi$ between the radial and azimuthal fluctuations.}
\label{fig:deltaP}
\eef
Fig.~\ref{fig:deltaP_ps} shows the expected level of period variation in case of a surface magnetic field of $1~\mathrm{kG}$ and varying fluctuation parameters $A_V$, $A_B$ and $A_P$ (see Sec.~\ref{sec:ang_mom}). Even with fluctuation amplitudes as high as $A_V = A_B = A_P = 0.5$ which is not observed in HR1099 \citep[see][]{Frasca2005}, the resulting binary period variation amplitudes are still off by two orders of magnitude, no physically reasonable parameter range yields period variations as high as those observed in the system ($9.0 \cdot 10^{-5}$, \citet{Frasca2005}). Given that the redistribution processes are almost entirely governed by convection (see Sec.~\ref{sec:analytical}), the only degree of freedom left to produce larger period variations is the convective velocity inside the star, e.g., via the mixing-length parameter $\alpha_\mathrm{ml}$. Reproduction of the observed modulation period of $35~\yr$ imposes an activity cycle length of twice that value as the model predicts a 2:1 relation between the observed binary modulation period and the activity cycle length. Therefore, if the observed period variations are energetically and mechanically feasible we would expect an activity cycle in HR1099 of roughly $70~\yr$ while a number of authors claim to have found cycle lengths between $14.1 \pm 0.3~\yr$ and $19.5 \pm 2~\yr$ \citep[see][and references therein]{Lanza2006HR1099, Muneer2010, Perdelwitz2018}.\\
\bef
\centering
\resizebox{\hsize}{!}{\includegraphics{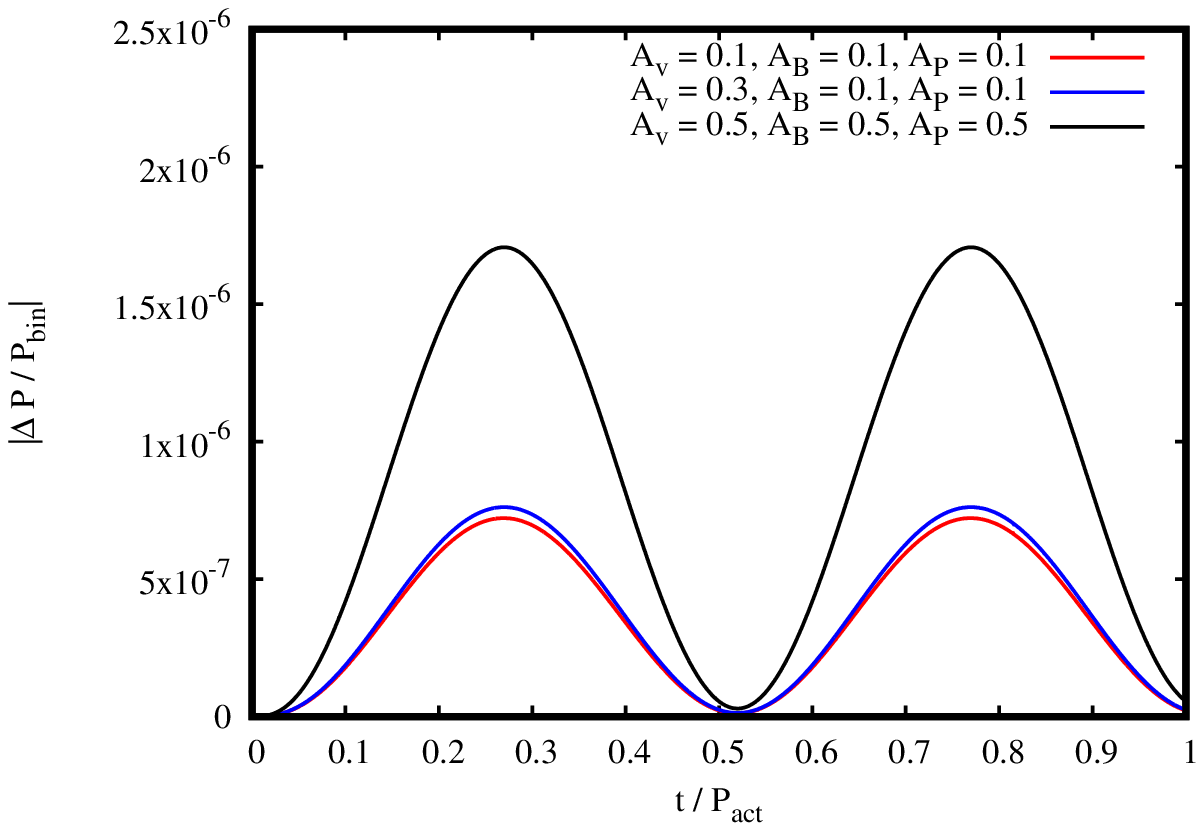}}
\caption{Binary period variation in HR1099 calculated for a surface magnetic field of $1~\mathrm{kG}$, a cycle length of $P_\mathrm{act} = 2 \, P_\mathrm{mod}$, no phase shift, and varying fluctuation parameters $A_V$, $A_B$ and $A_P$. In a conservative scenario with $A_P = A_V = A_B = 0.1$, the calculated binary period modulation amplitude is two orders of magnitude below the observed values ($\sim 10^{-4}$). Increasing the velocity field fluctuation amplitude to $A_V = 0.3$ at constant power dissipation $A_P = 0.1$ only results in a slight increase of $\Delta P / P$. Even for the rather extreme case of $A_P = A_V = A_B = 0.5$, the produced period variations are not in agreement with observations \citep[see][]{Frasca2005}.}
\label{fig:deltaP_ps}
\eef
\subsection{Active component mass parameter study}
\label{sec:mass_ps}
Having shown in the previous sections that RS CVns such as HR1099 are not expected to produce significant levels of period modulation, we expand our analysis to pre-cataclysmic PCEB systems \citep[see, e.g.][]{Zorotovic2013}, which have already been proposed as potential Applegate systems by \citet{Voelschow2016}.\\
Consider a tidally locked binary system with $a=1~\Rsol$, consisting of a typical White Dwarf primary with $M_\mathrm{WD} = 0.5~\Msol$ and a zero-age main sequence secondary with varying mass. Just as in the previous sections, we assume a constant magnetic field of $B_\mathrm{surf} = 1~\mathrm{kG}$ and fix the fluctuation parameters to $A_P = A_V = A_B = 0.1$ with a phase lag $\Delta \phi = 0$. For the modulation period, we assume $P_\mathrm{mod} = 15~\yr$. We varied the mass of the active component between $0.1~\Msol$ and $0.6~\Msol$ and calculated the amplitude of the orbital period variation using an EZAMS model of age $t = 0$. 
\bef
\centering
\resizebox{\hsize}{!}{\includegraphics{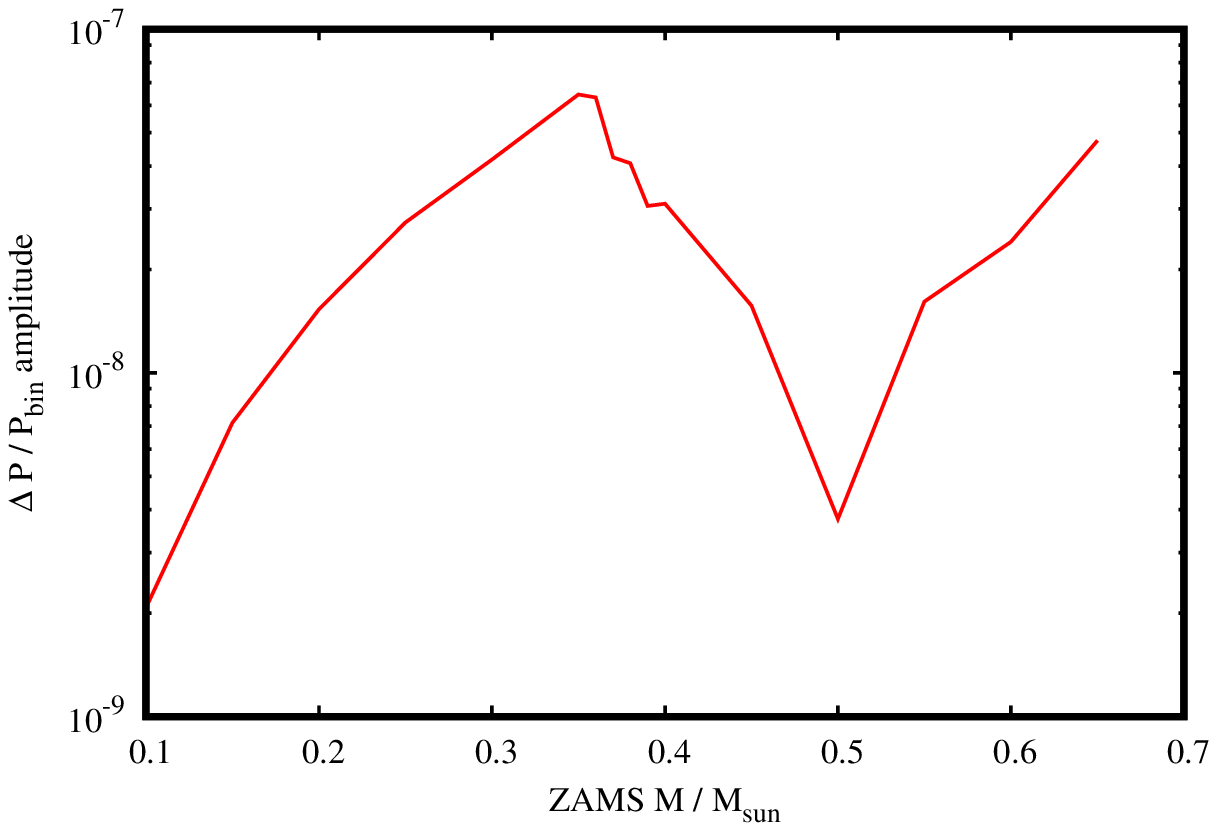}}
\caption{Amplitude of binary period variation calculated for a surface magnetic field of $1~\mathrm{kG}$, cycle length of $P_\mathrm{act} = 2 \, P_\mathrm{mod}$, magnetic field fluctuation amplitude of $A_B = 0.1$, velocity field fluctuations of $A_V = 0.1$ and varying active component mass. The noticeable decrease in period modulation at $0.37~\Msol$ coincides with the transition from fully-convective to radiative-core stars.}
\label{fig:mass_ps}
\eef
Fig.~\ref{fig:mass_ps} plots the expected period variation amplitude as a function of the active component mass. Between $0.1~\Msol$ and $0.36~\Msol$ we see a clear increasing trend towards higher binary period variation levels. For fluctuation parameters slightly above the fiducial $10~\%$ level, stars between $0.30~\Msol$ and $0.36~\Msol$ are expected to produce amplitudes $\simeq 10^{-7}$ which is a typical order of magnitude observed in PCEB systems \citep{Parsons2010, Zorotovic2013, Voelschow2016}, while stars outside of this range only support low levels of period variations $\lesssim 10^{-7}$. For all simulated systems, the peak angular velocity variations are $\lesssim 1~\%$.\\
The physical explanation for the peak around $M \simeq 0.35~\Msol$ is the emergence of a radiative core, which shifts the boundary of the convective zone outwards and reduces the total mass and angular momentum involved in the redistribution process. In Fig.~\ref{fig:convection}, we plot the bottom of the convective zone as function of mass. Stars with $M \gtrsim 0.37~\Msol$ develop a radiative core which reduces the domain available for angular momentum redistribution processes.
While the calculated binary period modulation amplitudes increase towards higher masses again, the upper end of the investigated mass-range may be affected by Roche-lobe overflow which introduces additional effects not considered in our model. Further, we remind the reader that fully-convective stars with $r_\mathrm{b}/R = 0$ are strictly speaking not covered by the regular Sturm-Liouville problem, but the structure files adopted in our calculations for stars up to $M = 0.36~\Msol$ have relative convection zone depths of $r_\mathrm{b}/R \simeq 0.01$.
\bef
\centering
\resizebox{\hsize}{!}{\includegraphics{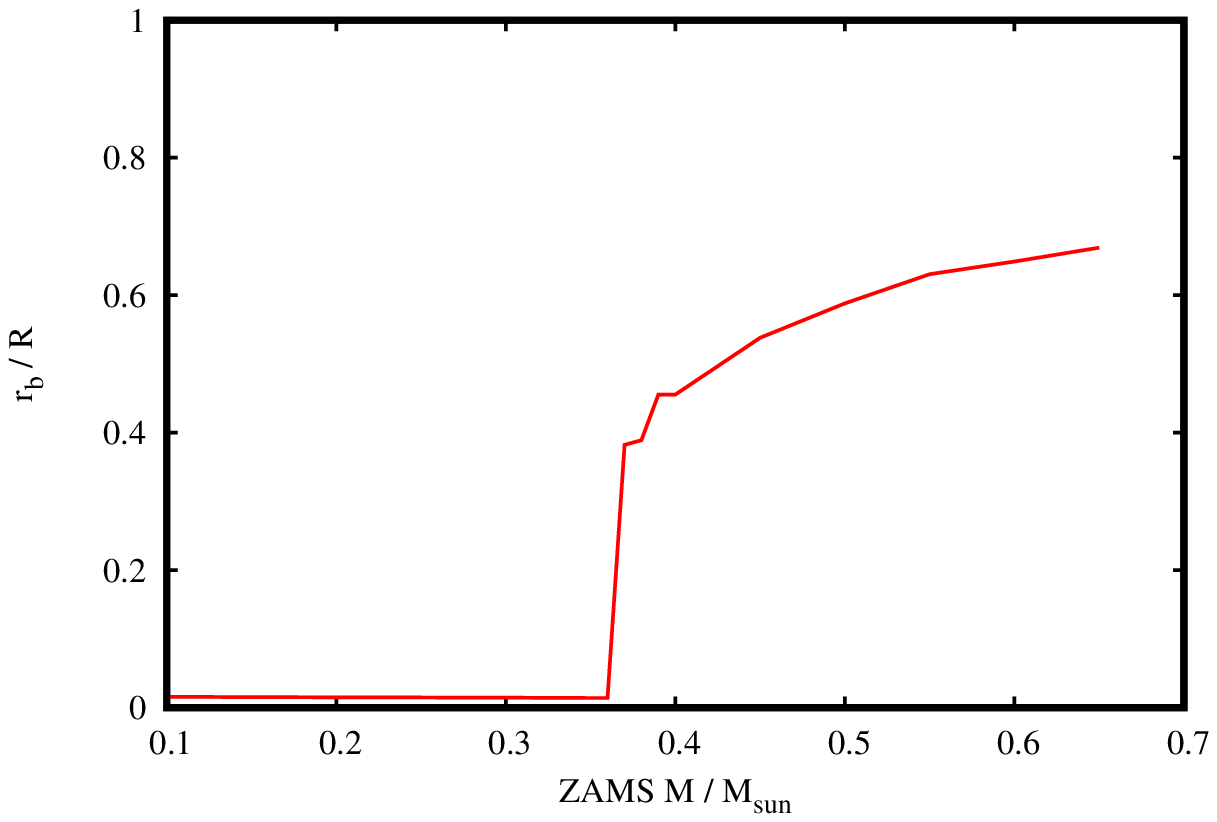}}
\caption{Bottom of the convection zone $r_\mathrm{b}$ as function of mass in our EZAMS stellar structure models. Stars up to $0.36~\Msol$ are fully convective with non-zero convection velocities down to the innermost grid points. For higher masses, EZAMS predicts the emergence of a radiative core which takes up increasing fractions of the central region and reduces the total mass participating in the angular momentum redistribution processes.}
\label{fig:convection}
\eef
\section{Conclusions and discussion}
We present a model that connects periodic fluctuations in the magnetic field and velocity field of the convection zone to periodic modulations of the quadrupole moment of the star, resulting in cyclic modulations of the binary period. Using the example of HR 1099, which has been discussed as one of the most promising Applegate candidates in the past, we calculated the angular momentum redistribution modes $\zeta _{nk}(r)$ and considered the angular equation to calculate the $\alpha _{nk}(t)$ and $\beta _{nk}(t)$ functions which are necessary to calculate the temporal evolution of the dissipated energy and the quadrupole moment, assuming that the star dedicates a fiducial fraction of $A_P = 0.1$ of its luminosity to drive the Applegate mechanism. The order up to which we calculate the eigenfunctions is determined by a simple soundwave travel time argument.\\
Our constant azimuthal field ansatz leads to a simple source function only controlled by the $\Lambda _{r\phi}$ and $M _{r\phi}$ components of the Reynolds and Maxwell tensors, where the former clearly dominates in the case of simple harmonic velocity and magnetic field fluctuations which we control via the $A_B$ and $A_V$ parameters with fiducial values of $0.1$.\\ 
In this model, harmonic oscillations of the magnetic field fluctuations and velocity field fluctuations on a timescale of the activity cycle $P_\mathrm{act}$ result in orbital period variations on a timescale of $P_\mathrm{mod} = 0.5 \, P_\mathrm{act}$. In other words, we expect that the activity cycle of any Applegate candidate satisfying our assumptions is double the observed binary modulation period whereas alternative models predict a different ration \citep[see, e.g.][]{Applegate1992, Lanza1998, Lanza2004}.\\
Furthermore, depending on the phase shift between the radial fluctuations and the azimuthal fluctuations, purely negative, positive or mixed quadrupole moment changes may be observed similar to findings by \citet{Ruediger2002}.\\
In line with previous work, we confirm that HR 1099 is not expected to produce period variations on the level observed today via the Applegate mechanism which would require unphysically high field fluctuation amplitudes, angular velocity changes and dissipated power. This is consistent with previous findings by \citet{Lanza2005} and \citet{Lanza2006} who did not explicitly consider the temporal evolution of the mechanism.\\
While HR 1099 in particular and RS CVn systems in general seem highly unlikely Applegate candidates, our extensive parameter study identified short-period PCEBs with active component masses between $0.30~\Msol$ and $0.36~\Msol$ as strongest Applegate candidates with expected period variation amplitudes similar to those typically observed in such systems which are between $10^{-7}$ and a few $10^{-6}$ \citep[see, e.g.][]{Zorotovic2013, Voelschow2016}. Consequently, we propose a careful re-analysis of PCEB systems with cyclic period variations to exclude the planetary hypothesis and find direct observational evidence that intrinsic effects such as quadrupole moment changes are at work.\\
To distinguish between Applegate modulations and period variations caused by planets via the light travel time effect \citep[see, e.g.][]{Pribulla2012}, long-term measurements are necessary to find correlations between activity indicators, luminosity variations due to photospheric temperature changes and the binary period modulation \citep[see, e.g.]{Applegate1992}. Changes in the rotational profile on the $1~\%$ level may lead to tiny deformations on a similar order of magnitude as the active component oscillates between an oblate and a prolate state which is also expected to correlate with the activity cycle \citep[see, e.g.]{Applegate1992, Lanza1998, Ruediger2002}. However, matters are complicated by spots and tidal deformations.\\
The model presented here is a significant step forward to understand the physics of the Applegate scenario. Nevertheless, it is still based on simplifying assumptions, such as a purely azimuthal magnetic field and the adoption of a mean-field framework. Also the Reynolds and Maxwell stress tensors do not self-consistently follow from a dynamo model, but we adopt a (physically motivated) prescription for the kinetic and magnetic fluctuations which is externally imposed. Similarly, this approach still neglects effects such as viscosity quenching due to fast rotation, which may reduce the dissipated power and allow for higher levels of period variation in rapidly rotating systems \citep[see][]{Lanza2006}. Furthermore, our model does not guarantee the hydrodynamical stability of the differential rotation profile that emerges during the oscillation, although the maximum relative deviations from the state of rigid rotation are on the $2~\%$ level in HR 1099 and well below the $1~\%$ level in our PCEB parameter study \citep[see][]{Knobloch1982}. Another limitation originates in the fact that we do not account for meridional circulation as well as the effect of energy advection inside the star. Finally, additional quadrupole moment changes may occur due to anisotropic Lorentz forces inside the star \citep{Lanza1998}.\\
While we cannot draw a final conclusion at this point, our framework nevertheless provides the most complete assessment of the Applegate mechanism so far, and shows that a consideration of physical processes allows to decide under which conditions the Applegate mechanism may be feasible. Our model already yields to testable predictions, suggesting that the Applegate mechanism may be favored for certain PCEB systems and disfavored in HR 1099. These predictions should be tested both on the observational side and through further refinement of the model to provide a clear picture where the Applegate mechanism is able to work.
\begin{acknowledgements}
We would like to express our gratitude to the anonymous referee for carefully proof-reading our manuscript and providing us with valuable advice that improved the quality of our work. DRGS thanks for funding via Fondecyt regular (project code 1161247), the ''Concurso Proyectos Internacionales de Investigaci\'on, Convocatoria 2015'' (project code PII20150171) and the BASAL Centro de Astrof\'isica y Tecnolog\'ias Afines (CATA) PFB-06/2007.
\end{acknowledgements}
\bibliography{astro.bib}
\bibliographystyle{aa}
\end{document}